\documentclass{article}
\usepackage[english]{babel}
\usepackage{float}
\usepackage{booktabs}
\usepackage{graphicx}
\usepackage{dcolumn}
\usepackage{bm}
\usepackage{cite}
\usepackage{amsthm} 
\DeclareUnicodeCharacter{2212}{$-$}

\usepackage[letterpaper,top=2cm,bottom=2cm,left=3cm,right=3cm,marginparwidth=1.75cm]{geometry}

\usepackage{amsmath}
\usepackage{graphicx}
\usepackage[colorlinks=true, allcolors=blue]{hyperref}
\usepackage{authblk}
\usepackage{tabularx}

\makeatletter
\renewcommand{\AB@authnote}[1]{}
\renewcommand{\AB@affilnote}[1]{}
\makeatother

\title{The golden cage of science: first-last author prestige gap buys impact over disruption}
\author{Chi Zhang}
\author{Yilin Bi}
\author[$\ast$]{Tao Zhou}
\affil{CompleX Lab, School of Computer Science and Engineering, University of Electronic Science and Technology of China, Chengdu 610054, China}
\affil[$\ast$]{To whom correspondence should be addressed: \href{zhutou@ustc.edu}{zhutou@ustc.edu}}

\begin{document}
\maketitle

\begin{abstract}
First and last authors make critical but distinct contributions in the majority of scientific papers. First authors tend to conduct research and analysis, whereas last authors are usually responsible for setting research directions and providing resources and supervision. The first-last author prestige gap (FL gap, $\Delta H$) is defined as the H-index difference between the last and first authors prior to the focal paper's publication. The FL gap serves as a measure of the prestige asymmetry between the two dominant collaborators. We find that larger FL gaps in either direction are associated with higher impact and lower disruption. This trade-off is robust across team sizes and intensities of prior collaboration and can not be explained by first and last authors' separate prestige effects or their academic age gap. Further analysis shows that variation in last author prestige is associated with larger differences in both the article's impact and disruption than variation in first author prestige. We refer to this coexistence of higher impact and lower disruption related to the FL gap as the ``golden cage of science'', which reflects a common collaboration structure that statistically produces highly cited but less disruptive work. These findings suggest the first-last author prestige gap as an important indicator for examining how research teams balance senior guidance with exploratory autonomy.
\end{abstract}


\vspace {10mm}
Scientific progress depends on both consolidating knowledge within existing paradigms and disruptive innovation \cite{Kuhn1962,Funk2017}. Over the past half-century, research outputs have become less disruptive despite substantial growth in publication volume \cite{Park2023}. Generating novel ideas has become increasingly difficult \cite{Bloom2020}, and the likelihood of going beyond established conceptual frameworks has decreased \cite{Rzhetsky2015,Chu2021}. Team collaboration has largely replaced individual authorship as the dominant mode of scientific research \cite{Price1963,Wuchty2007,Zeng2017,Fortunato2018}. Recent studies have therefore examined associations between scientific disruption and team characteristics such as team size \cite{Wu2019}, team freshness \cite{Zeng2021,Yang2025}, diversity among team members \cite{Hofstra2020,Zhang2024,Kamal2026} and geographical distance between collaborators \cite{Lin2023}.

These studies link team composition and collaboration patterns to scientific disruption, but rarely distinguish the contributions of individual members. In particular, how the characteristics of dominant contributors relate to scientific disruption remains unclear. In papers with multiple authors, author order often reflects authors' roles and contributions \cite{VanDijk2014,Lariviere2016,Haeussler2020}. In the majority of scientific papers, first and last authors commonly make substantial contributions \cite{Lariviere2016}, although their roles differ. First authors typically serve as the primary collaborator, conducting research and leading the analysis. Last authors often set research directions, provide essential scientific problems, offer necessary resource, and review the work \cite{Baerlocher2007,Tscharntke2007,Wren2007}. Last authors are usually more senior and have greater academic prestige than first authors. A common combination is a student as the first author and her/his supervisor as the last author. In practice, other combinations of first and last authors' prestige are also possible. For instance, two senior researchers with similar prestige may occupy these positions, or the more senior researcher may conduct the work and be listed as the first author. We consider the prestige gap between first and last authors an important measure of collaboration patterns, one that has received little attention in previous studies. Specifically, we quantify this gap as the last author's H-index minus the first author's, both calculated before the focal paper's publication. We then examine its relationship to the corresponding article's impact and disruption.

We analyze publication and citation records from 1990 to 2020 in the Microsoft Academic Graph (MAG) \cite{Sinha2015}. To allow sufficient time for citations to accumulate, our analysis focuses on journal articles published between 1990 and 2010. With citation records through 2020, each focal paper therefore has a citation window of at least ten years, which reduces the instability and bias related to limited citation accumulation \cite{Bornmann2019}. We exclude journal articles with no recorded citations from the analysis to ensure the validity of metrics. This paper focuses on eight disciplines with established authorship conventions: medicine, chemistry, physics, computer science, sociology, economics, political science, and history. The MAG dataset contains 7,996,043 cited journal articles published in these disciplines between 1990 and 2010. After excluding single-author articles and articles with alphabetically ordered author lists, our sample contains $n = 5,305,920$ journal articles. The MAG dataset provides unique author identifiers after author disambiguation, allowing publication records to be linked to individual researchers.

\subsection*{\large Prestige gap buys impact over disruption}
\begin{figure}[!t]
    \centering
    \includegraphics[width=1\linewidth]{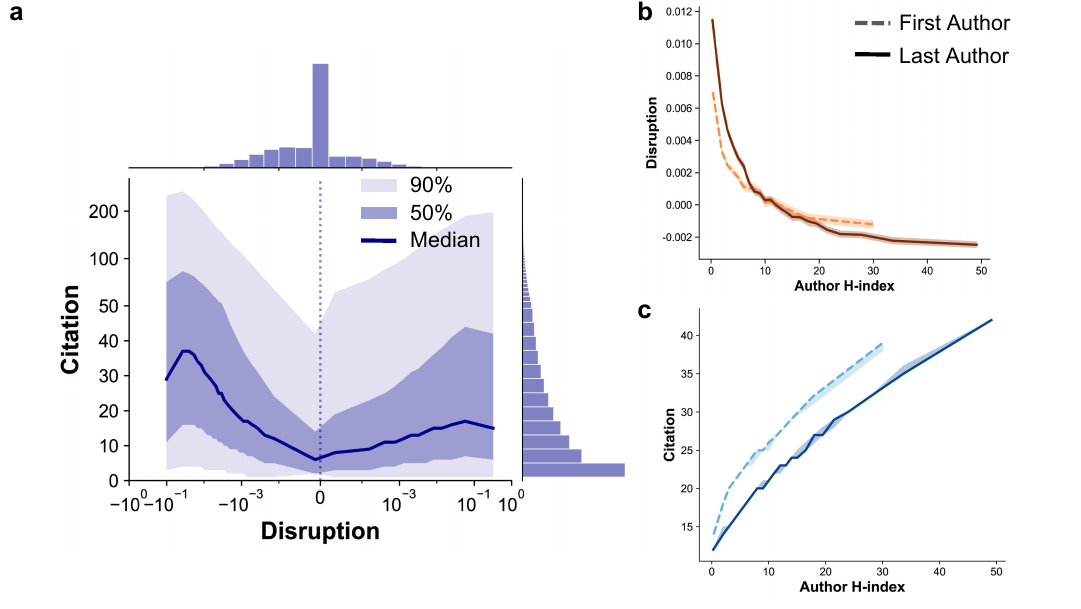}
    \caption{\textbf{Distributions of disruption and impact and their relationships with first and last author H-index. }\textbf{a, }Joint distribution of disruption (measured by the disruption index) and impact (measured by cumulative citations). The line shows the median citation at each disruption level. Dark and light shading zones represent the 25th--75th and 5th--95th percentile intervals, respectively. Median citation varies nonlinearly with mean disruption. The wide percentile intervals indicate large variation in citations across disruption levels. \textbf{b,c, }Trends in disruption and impact as first and last author H-indices change. Dashed and solid curves represent first and last authors respectively. As H-index increases, the mean disruption significantly decreases from the 88th to the 47th percentile for first authors and from the 91st to the 39th percentile for last authors (\textbf{b}). Median citation increases from the 43rd to the 72nd percentile for first authors and from the 40th to the 75th percentile for last authors (\textbf{c}). Bootstrapped 90\% confidence intervals are shown as shaded bands in \textbf{b,c}.}
    \label{fig1}
\end{figure}

We assess each article in terms of impact and disruption. Impact is measured by cumulative citations received through 2020. Disruption is quantified using the disruption index $D$. The disruption index was originally developed to distinguish between consolidating and destabilizing patent technologies \cite{Funk2017} and was later applied to quantify disruption of scientific papers \cite{Wu2019}. For a focal paper, the disruption index $D$ is defined as 
\begin{equation}
D = \frac{N_i - N_j}{N_i + N_j + N_k},
\end{equation}
where $N_i$ is the number of subsequent articles that cite the focal paper but none of its references, $N_j$ counts subsequent articles that cite both the focal paper and at least one of its references, and $N_k$ counts subsequent articles that cite at least one of the focal paper's references but not the focal paper. $D$ ranges from -1 to 1. $D = 1$ indicates the most disruptive article, where all subsequent citing articles cite only the focal paper. Conversely, $D = -1$ indicates the most consolidating article, where all subsequent citing articles cite both the focal paper and its references. The distributions of citation counts and $D$ values, together with their relationship, are shown in Figure~\ref{fig1}a. Citation counts are concentrated at low values and 79\% of articles received less than 50 citations. Disruption index values gather around 0, with approximately 30\% of articles falling within $[-0.001,0.001]$. The overall correlation of the two metrics is low (Pearson's $r = 0.014$), and the joint distributions across different thresholds suggest that there is no simple correlation between citation counts and disruption (Figure~S1). These results support evaluating research outputs in terms of both impact and disruption.

We next examine how author prestige relates to article impact and disruption. The H-index is used as a measure of accumulated prestige at a given time\cite{Bornmann2005,Hirsch2005,Cronin2006,VanRaan2006}. An author has an H-index of $h$ if at least $h$ of her/his articles have each received at least $h$ citations, but fewer than $h+1$ articles have each received at least $h+1$ citations \cite{Hirsch2005}. To account for changes in prestige over time, we calculate each author's historical H-index. For a focal paper published in year $t$, only publication and citation records before year $t$ are included in the calculation. For first and last authors, who generally play key roles in the research, an increase in prestige corresponds to a sustained increase in citation and a gradual decline in disruption (Figure~\ref{fig1}b and Figure~\ref{fig1}c). Although the trends are similar for both author positions, the curves remain distinct.

\begin{figure}[!t]
    \centering
    \includegraphics[width=\linewidth]{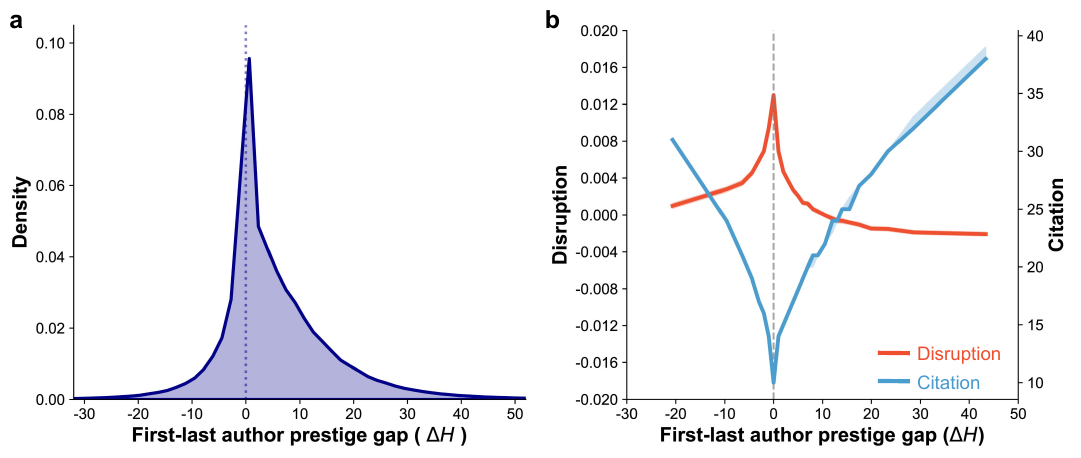}
    \caption{\textbf{Distribution of the FL gap and the trade-off between impact and disruption. } \textbf{a, }The distribution of $\Delta H$ is right-skewed, with the majority of observations near or above zero. Values near 0 indicates similar first and last author prestige, and positive values indicate that the last author has higher prestige. 
    \textbf{b, }The trade-off related to $\Delta H$. Mean disruption (red curve, indexed by left $y$ axis) shows a significant inverted V-shaped relationship with the FL gap, whereas median citation (blue curve, indexed by right $y$ axis) shows a V-shaped relationship. Both curves show a sharp turning point at $\Delta H = 0$. At this point, mean disruption reaches its peak of approximately 0.013, while median citation reaches its minimum of approximately 10. As the FL gap widens in either direction, mean disruption declines and median citation increases. These changes in the two directions are asymmetric. For $\Delta H > 0$, the mean disruption reaches a lower minimum of approximately −0.002, whereas median citation reaches a higher maximum of approximately 38. Bootstrapped 90\% confidence intervals are shown as shaded bands.}
    \label{fig2}
\end{figure}

We use the first-last author prestige gap (FL gap) to quantify the relative prestige of the two dominant authors. It is defined as
\begin{equation}
\Delta H = H_{last} - H_{first},
\end{equation}
where $H_{last}$ and $H_{first}$ are the last and first authors' H-index values before the focal paper's publication. The distribution of $\Delta H$ is right-skewed, and 77\% of articles have $\Delta H \geq 0$ (Figure~\ref{fig2}a). Thus, last author prestige is at least as high as first author prestige in the majority of articles.

The FL gap reveals a clear trade-off between impact and disruption. Impact shows a V-shaped relationship with the FL gap, reaching its minimum when first and last authors have similar prestige ($\Delta H \approx 0$) (Figure~\ref{fig2}b). Articles with larger gaps in either direction receive more citations, whether the higher-prestige author occupies the first ($\Delta H < 0$) or the last ($\Delta H > 0$) position. Conversely, disruption shows an inverted V-shaped pattern (Figure~\ref{fig2}b), declining as the prestige gap widens. To assess whether this trade-off could be explained by the first and last authors' separate prestige effects, we construct a null model (see Methods). The result shows that even an optimally weighted linear combination of these effects cannot fully reproduce the observed pattern (Figure~S2). For articles with larger FL gaps ($\lvert \Delta H \rvert > 5$), observed disruption is systematically lower and impact higher than predictions by the null model.

\subsection*{\large Last author has higher influence}
\begin{figure}[!t]
    \centering
    \includegraphics[width=1\linewidth]{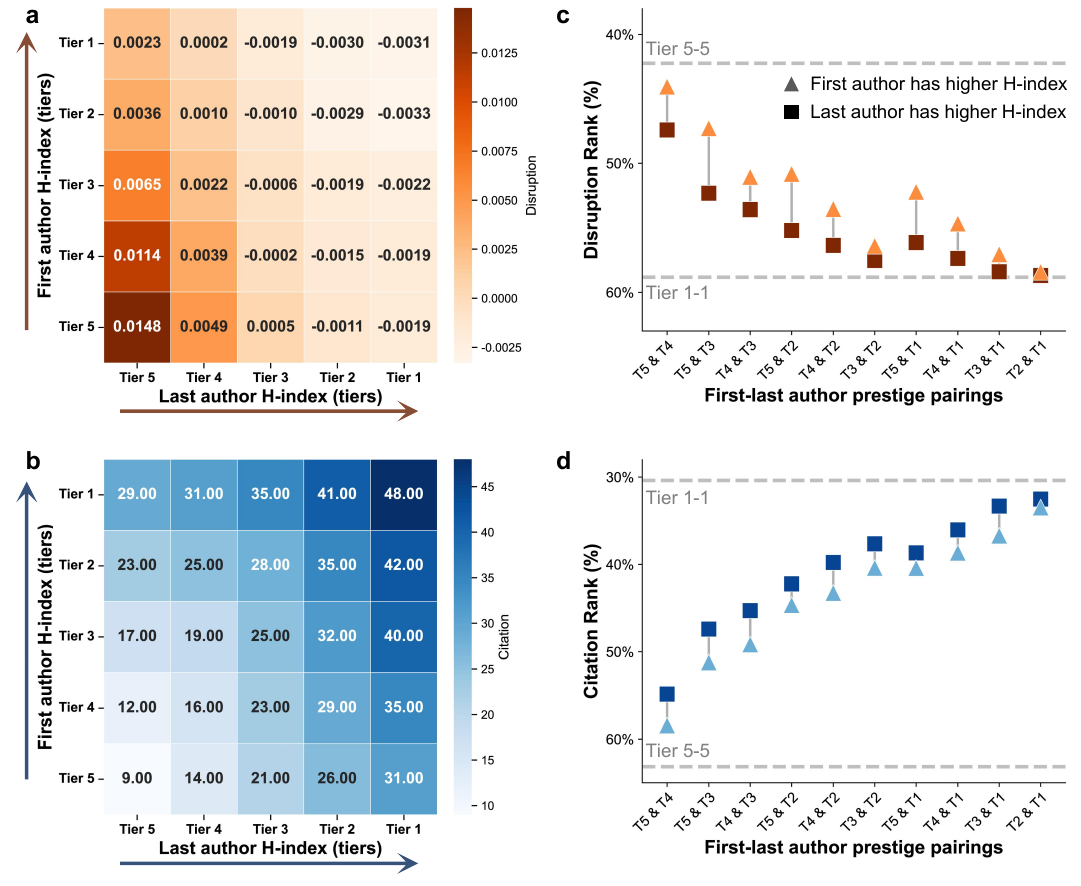}
    \caption{\textbf{Disruption and impact across first-last author prestige pairings. }First and last authors are separately classified into five H-index tiers, with Tier 1 representing the highest prestige and Tier 5 the lowest. \textbf{a,b, }Mean disruption (\textbf{a}) and median citation (\textbf{b}) across the 25 tier combinations. Articles with both authors in Tier 5 (Tier~5-5) have the highest mean disruption (0.0148) and the lowest median citation (9.00). By contrast, the mean disruption of articles with both authors in Tier 1 (Tier~1-1) drops to a negative value of −0.0031, and the median citation rises to 48.00, more than five times that of Tier~5-5. With the other author's tier fixed, shifts in last author tier along the horizontal axis correspond to larger differences in both outcomes than shifts in first author tier. \textbf{c,d, }Comparisons within each different-tier pairing, with the higher-prestige author occupying the first position (triangles) or last position (squares). The horizontal axis identifies the first-last author prestige pairings. The vertical axes show mean disruption rank (\textbf{c}) and mean citation rank (\textbf{d}). Ranks range from 0\%, the best rank, to 100\%, the worst rank. Horizontal dashed gray lines indicate the Tier~5-5 and Tier~1-1 benchmarks. Across all pairings, articles with the higher-prestige researcher as last author show worse disruption ranks (\textbf{c}) and better citation ranks (\textbf{d}).
    }
    \label{fig3}
\end{figure}

First and last authors are separately classified into five prestige tiers based on their H-index values to compare article outcomes across different prestige combinations. To account for the long-tailed prestige distribution, narrower percentile intervals at the upper end are used. Tier 1 to 5 correspond to the top 5\%, 5\%--10\%, 10\%--40\%, 40\%--70\% and 70\%--100\%, respectively. Comparisons across the resulting 25 prestige combinations reveal differences in article outcomes. The Tier~5-5 combination, in which both first and last authors are in the lowest tier, has the highest mean disruption and the lowest median citation. The Tier~1-1 combination, in which both authors are in the highest tier, shows the opposite pattern, combining low disruption with high citation (Figure~\ref{fig3}a and Figure~\ref{fig3}b).

Given the last author tier, a higher prestige tier for the first author corresponds to higher citation and lower disruption. The same pattern holds when first author tier is fixed and last author prestige tier varies. However, the magnitude of these differences depends on author position. As shown in Figure~\ref{fig3}a and Figure~\ref{fig3}b, when the other author's tier is fixed, differences in both impact and disruption are larger across last author tiers than across first author tiers. This finding is consistent with the separate relationships between first and last author prestige and article outcomes (Figure~\ref{fig1}b and Figure~\ref{fig1}c).

Within each pairing of different prestige tiers, articles with the last author in the higher prestige tier have worse mean disruption ranks but better mean citation ranks than those with the first author in the higher prestige tier (Figure~\ref{fig3}c, Figure~\ref{fig3}d, and Figure~S3). This pattern holds across all different-tier pairings. With articles at $\Delta H \approx 0$ as the baseline for comparison, for comparable $|\Delta H|$ values, the conventional combination with $\Delta H>0$ corresponds to larger citation gains and disruption declines than the reverse combination with $\Delta H<0$. This asymmetry is consistent with Figure~\ref{fig2}b, in which the positive region reaches a higher maximum citation and a lower minimum disruption than the negative region. The shifts in citation and disruption ranks could differ substantially in magnitude within some pairings (Figure~\ref{fig3}c, Figure~\ref{fig3}d, and Figure~S3). For example, within the T5 \& T1 pairing, the more common combination places the Tier-1 author last (Tier~5-1). Compared with articles in Tier~5-1, those with the Tier-1 author first and the Tier-5 author last (Tier~1-5) have a mean citation rank that is 1.74\% worse, while their mean disruption rank is 3.92\% better.

\subsection*{\large Robustness and sensitivity analysis}
\begin{figure}[!t]
    \centering
    \includegraphics[width=1\linewidth]{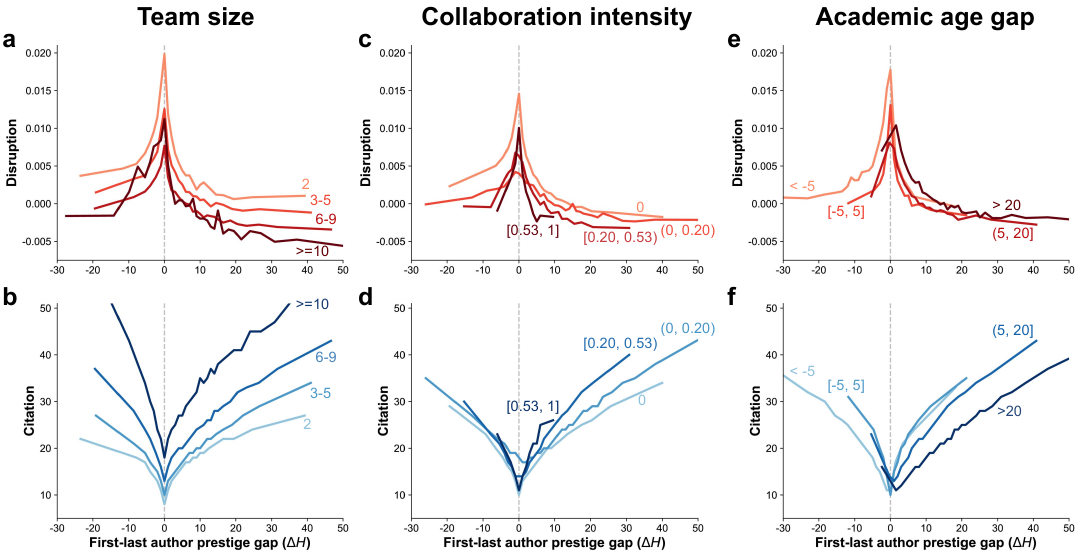}
    \caption{\textbf{Robustness of the trade-off across team size, collaboration intensity and academic age gap. }\textbf{a,b, }In each of the four team size groups, the inverted V-shaped relationship for disruption and V-shaped relationship for impact persist, despite differences in baselines related to team size. \textbf{c,d, }Prior collaboration intensity does not qualitatively change the observed relationships, although higher intensity corresponds to a lower disruption baseline. The highest mean disruption occurs among articles whose first and last authors have similar prestige and have not co-authorized articles before. \textbf{e,f, }Curves representing different academic age gaps overlap extensively, retaining the V-shaped and inverted V-shaped relationships. The total sample size for each panel is $n = 5,305,920$ articles. Details about variable definitions, grouping criteria and group sample sizes are provided in Methods.}
    \label{fig4}
\end{figure}

The trade-off between impact and disruption remains robust across different collaboration contexts. Although article outcomes vary with team size \cite{Wu2019}, their relationships with the FL gap remain consistent. Across two-author, small, medium-sized and large teams, disruption is still highest when first and last authors have similar prestige ($\Delta H \approx 0$) and also declines as the gap widens (Figure~\ref{fig4}a). Impact follows the opposite pattern (Figure~\ref{fig4}b). Notably, article outcomes vary with both team size and the FL gap. When first and last authors have similar prestige, large teams can produce articles with disruption comparable to that of small teams. Even two-author teams will lose the potential disruption advantage of small teams when the FL gap is large. Larger teams also maintain higher citation across the range of FL gaps. Thus, alongside differences in impact baselines across team sizes, the FL gap further distinguishes disruption within each size group.

We further examine whether prior collaboration intensity and academic age gap account for the trade-off. We quantify collaboration intensity using the number of articles previously coauthored by the first and last authors, normalized by the geometric mean of their prior publication counts (see Methods). The inverted V-shaped relationship between the FL gap and disruption persists across different collaboration intensity levels. However, higher collaboration intensity corresponds to a lower disruption baseline, potentially reflecting greater knowledge homogeneity. All collaboration intensity levels nevertheless show the same trade-off between impact and disruption (Figure~\ref{fig4}c and Figure~\ref{fig4}d). To analyze generational differences, we define the academic age gap as the difference between the dates of the first and last authors' first publications. We group articles by the academic age gap (see Methods), and the curves largely overlap across groups (Figure~\ref{fig4}e and Figure~\ref{fig4}f), suggesting that academic age gaps add little explanatory information once the FL gap is considered. The trade-off therefore cannot be readily explained by generational differences. The trade-off also persists when articles are grouped separately by citation and disruption percentiles (Figure~S4), indicating that the trade-off does not arise solely from aggregating articles with different outcome levels. This phenomenon remains robust across publication years (Figure~S5) and disciplines (Figure~S6), although its magnitude and details vary across disciplines. We also distinguish original research articles from reviews using natural language processing techniques (see Methods). Both article types show qualitatively similar patterns (Figure~S7). Together, these analyses indicate that the FL gap trade-off extends across outcome levels, publication years, disciplines and article types. It cannot be explained simply by differences in team size, prior collaboration intensity or academic age gap.

To assess the sensitivity of this trade-off to the definition of the disruption index, we evaluate index variants that apply citation thresholds, reduce noise, exclude self-citations or account for citation network connectivity (Table~S1) \cite{Wu2019,Bornmann2020,Yang2024}. Their relationships with the FL gap all remain inverted V-shaped (Figure~S8a). Heatmaps of the 25 first-last author prestige combinations also show similar patterns for all different disruption variants examined (Figure~S8b, Figure~S8c, Figure~S8d, Figure~S8e, Figure~S8f, and Figure~S8g).

\section*{\large Discussion}
Our findings identify the prestige gap between first and last authors as an important dimension for understanding differences in team research outputs. As the FL gap expands, articles tend to exhibit higher impact and lower disruption, reflecting a consistent trade-off. This trade-off persists across team sizes, prior collaboration intensities and academic age gaps, and remains robust to alternative disruption metrics. The phenomenon reflect both the authors' individual academic prestige and the interactions and dependencies arising during team collaboration. Cumulative advantage can help high-prestige researchers sustain their visibility and access to resources \cite{Merton1968,Price1976,Petersen2014}. They can draw on their accumulated experience, resources and broad academic perspectives to enhance the team's visibility \cite{Sekara2018}. Meanwhile, hierarchical authority structures within teams can constrain the generation of innovative ideas \cite{Keum2017}. These effects are not solely determined by one party's prestige and should be understood in terms of the interactions and dependencies between first and last authors. By exchanging data, resources and expertise, scientific collaborators can thus undertake a wider range of research \cite{Morrison2017}. The resulting dependence is shaped by both collaborators' research needs and access to alternative resources, rather than by one collaborator's academic attributes alone \cite{Emerson1962}. We therefore conjecture that research directions and strategies may align more closely with one author's research agenda when that author is the main source of academic resources and visibility and the other has limited access to alternative resources. When the decision-making authority rests primarily with the author who holds more academic capital, the team's research strategy is more likely to shift towards consolidation and advancement of established frameworks \cite{March1991,Foster2015}. Prestige-based endorsement, hierarchical team structures and resource dependence may thus jointly contribute to the ``golden cage of science'', in which high impact coexists with low disruption.

These findings offer a noteworthy perspective on the organization and design of research teams. The key question may be how to balance senior guidance with exploratory autonomy by adjusting team composition and the division of work, rather than whether to adopt hierarchical or flat team structures. This balance could support both the steady advancement of established knowledge and disruptive innovation. For example, within the Tier 5 \& Tier 1 pairing, mean disruption rank is about 4\% better with the Tier-1 researcher as first rather than last author. This suggests that, in addition to setting research directions, senior researchers may draw on their experience to identify opportunities for innovation through hands-on research. Such involvement could contribute to more disruptive work.

Our study has several limitations. Our observational data do not allow us to isolate the causal effects of prestige structure from those of team formation. Existing teams may select research tasks and strategies based on their strengths, while task requirements may also shape team composition. Our findings are therefore correlational and cannot determine how much of the observed trade-off is attributable to the team's prestige structure itself. Furthermore, direct measures of resource support, direction-setting and exploratory autonomy are not available. Without these data, we cannot directly test the roles of these factors in the proposed organizational mechanisms. Future studies could track team formation and each author's specific contributions. They could compare research outputs from teams formed before task specification with those from teams assembled for predefined tasks, at comparable FL gaps. By combining data on funding allocations and individual contributions, further research could clarify the role of research autonomy in teams with comparable prestige gaps.

\section*{Methods}
\subsection*{Null Model}
We construct a null model to assess whether the observed trade-off could be reproduced by combining predictions based separately on first and last author prestige. The model represents a focal paper's predicted outcome as a weighted linear combination of the outcomes corresponding to first and last authors' H-index values. It therefore includes no interaction term between those two authors' prestige. If the null model could fully reproduce the trade-off, the FL gap would not be an informative parameter.

For focal paper $i$, let $Y$ denote either disruption $D$ or impact $C$, with $Y\in\{D,C\}$. The null model prediction is given by
\begin{equation}
Y_{null,i}=(1-\beta_Y)Y_{first,i}+\beta_Y Y_{last,i}.
\end{equation}
where $Y_{first,i}$ and $Y_{last,i}$ are predictions based on the first and last authors' H-index values, respectively. For $Y = D$, the predictions are conditional means. For $Y = C$, they are conditional medians. Problems of estimating the optimal coefficients $\beta_Y^*$ are reduced to ordinary least squares (OLS) problems with mean squared error (MSE) shown as below:
\begin{equation}
MSE(\beta_Y) = \frac{1}{N} \sum_{i=1}^{N} [(Y_{real, i} - Y_{first, i}) - \beta_Y (Y_{last, i} - Y_{first, i})]^2,
\end{equation}
where $Y_{real, i}$ is the observed outcome of focal paper $i$. Differentiating the MSE with respect to $\beta_Y$ and setting the derivative to zero give:
\begin{equation}
\beta_Y^* = \frac{\sum (Y_{real, i} - Y_{first, i}) (Y_{last, i} - Y_{first, i})}{\sum (Y_{last, i} - Y_{first, i})^2}.
\end{equation}
The optimal weights are $\beta_D^* = 0.8257$ for disruption ($R^2 = 0.8076$), and $\beta_C^* = 0.6463$ for impact ($R^2 = 0.7177$). These $R^2$ values indicate that the null model captures much of the overall pattern. However, the curves predicted using these optimal weights $\beta_D^*$ and $\beta_C^*$ still systematically deviate from the observed ones (Figure~S2). Such deviations suggest that the relationships between first and last authors could influence the article's performance, beyond the separate prestige effects represented in this linear model.

\subsection*{Control Variables}
We examine a number of variables to assess the robustness of the observed patterns. Publication year and discipline are directly available in the MAG dataset. Other considered variables are described below.

\textbf{\textit{Team size. }}The median team size is four, with an interquartile range of three to six. The grouping considers both the sample distribution and common team sizes in practice. Articles are grouped into four categories: 2, 3–5, 6–9 and $\geq 10$ authors. These four groups contain $707,266$, $2,986,211$, $1,323,715$ and $288,728$ articles, respectively.

\textbf{\textit{Collaboration intensity. }}To account for potential cognitive inertia from prior collaboration, we define the collaboration intensity ($I$) between first and last authors $A$ and $B$ as
\begin{equation}
    I_{AB} = \begin{cases}
        \frac{N_{AB}}{\sqrt{N_{A} N_{B}}}, & N_{A} N_{B} > 0, \\[0.5em]
        0, & N_{A} N_{B} = 0.
    \end{cases}
\end{equation}
Here, $N_{AB}$ is the number of articles previously coauthored by $A$ and $B$, and $N_{i}$ is the number of articles authored by $i \in \{A, B\}$ prior to the focal paper. Collaboration intensity ($I$) ranges from 0 to 1. A value of zero indicates no prior coauthored publications while values close to one indicate extensive overlap in the authors' prior publication records. Articles are assigned to four groups: $I = 0, 0 < I < 0.20, 0.20 \leq I < 0.53, 0.53 \leq I \leq 1$, containing $2,869,745$, $1,212,687$, $979,499$ and $243,989$ articles, respectively. Articles with $I = 0$ account for 54\% of the sample and are therefore treated as a single group to distinguish author pairs with and without prior collaboration. The values 0.20 and 0.53 correspond to the 50th and 90th percentiles of nonzero $I$ values, respectively.

\textbf{\textit{Academic age gap. }}To ensure that the observed FL gap patterns are not explained solely by generational differences among researchers, we examine the academic age gap between first and last authors. We calculate the age gap by subtracting the last author's first publication year from the first author's, based on publication records in the MAG dataset. For example, if the first author's first article is published in 2009 and the last author's in 1994, then the gap is $2009-1994=15$ years. The grouping considers typical timelines for scientific training and usual career development, with sufficient articles in each group. Articles are grouped by academic age gap: $<-5$, $[-5,5]$, $(5,20]$ and $>20$ years, containing $829,654$, $1,695,946$, $1,892,750$ and $887,570$ articles, respectively.

\textbf{\textit{Article type. }}Review articles synthesize and consolidate existing research and are therefore expected to exhibit low disruption. They differ from original research articles in publication purposes. Without article type labels in the MAG dataset, we distinguish reviews from original research articles using a natural language processing (NLP) model based on their titles and abstracts. The SciBERT model \cite{Beltagy2019} is fine-tuned using articles with explicit type labels from PubMed, supplemented by MAG records whose article types can be clearly identified. The fine-tuned model achieves a precision of 90.45\% and an F1-score of 92.79\%. We apply the model to our analytical sample ($n = 5,305,920$) and identify $n_{o} = 3,989,519$ original research articles (75.2\%) and $n_{r} = 1,316,401$ reviews (24.8\%). Review articles have a mean disruption of $−0.0033$ and a median citation of 27; the corresponding values for original research articles are $0.0068$ and 16.

\bibliographystyle{plain}

\end{document}